\documentclass[preprint]{aastex}

\shorttitle{Giant H~II Regions. IV. NGC3576}
\shortauthors{Figuer\^ edo et al.}

\begin{document}

\title{The Stellar Content of Obscured Galactic Giant H~II Regions IV.: NGC3576}

\author{E. Figuer\^ edo}
\affil{IAG-USP, R. do Mat\~ao 1226, 05508-900, S\~ ao Paulo, Brazil} 
\email{lys@astro.iag.usp.br}

\author{R. D. Blum\altaffilmark{1}}
\affil{Cerro Tololo Interamerican Observatory, Casilla 603, La Serena, Chile} 
\email{rblum@noao.edu}

\author{A. Damineli\altaffilmark{1}}
\affil{IAG-USP, R. do Mat\~ao 1226, 05508-900, S\~ ao Paulo, Brazil}
\email{damineli@astro.iag.usp.br}

\and

\author{P. S. Conti}
\affil{JILA, University of Colorado \\ Campus Box 440, Boulder, CO, 80309}
\email{pconti@jila.colorado.edu}

\altaffiltext{1}{Visiting Astronomer, Cerro Tololo Inter-American Observatory,
National Optical Astronomy Observatories, which is operated by Associated
Universities for Research in Astronomy, Inc., under cooperative agreement with
the National Science Foundation.}

\begin{abstract}
We present deep, high angular resolution near--infrared images of the
obscured Galactic Giant H~II region NGC3576. Our images reach objects
to $\sim 3M_\odot$.  We collected high signal--to--noise $K-$band
spectra of eight of the brightest objects, some of which are affected
by excess emission and some which follow a normal interstellar
reddening law. None of them displayed photospheric features typical of
massive OB type stars. This indicates that they are still enshrouded
in their natal cocoons. The $K-$band brightest source (NGC3576\#48)
shows CO 2.3 $\mu$m bandhead emission, and three others have the same
CO feature in absorption.  Three sources display spatially unresolved
$H_2$ emission, suggesting dense shocked regions close to the
stars. We conclude that the remarkable object NGC3576\#48 is an
early--B/late--O star surrounded by a thick circumstellar
disk/envelope. A number of other relatively bright cluster members
also display excess emission in the $K-$band, indicative of
reprocessing disks around massive stars (YSOs). Such emission appears
common in other Galactic Giant H~II regions we have surveyed.  The IMF
slope of the cluster, $\Gamma = -1.62$, is consistent with Salpeter's
distribution and similar to what has been observed in the Magellanic
Cloud clusters and in the periphery of our Galaxy.

\end{abstract}


\section{Introduction}

Massive stars have a strong impact on the evolution of
galaxies. O--type stars and their descendants, the Wolf--Rayet stars,
are the main source of UV photons, mass, energy and momentum to the
interstellar medium. They play the main role in the ionization of the
interstellar medium and dust heating. The Milky Way is the best place
to access, simultaneously, massive stellar populations and their
impact on the surrounding gas and dust. The sun's position in the
Galactic plane, however, produces a heavy obscuration in the optical
window ($A_V \approx 20 - 40$ mag) toward the inner Galaxy, where
massive star formation activity is the greatest.  Shifting to longer
wavelengths, the perspective is much better, especially in the near
infrared, because these wavelengths are long enough to lessen the
effect of interstellar extinction ($A_K \approx 2 - 4$ mag) and still
short enough to probe the stellar photospheric features of massive
stars \citep{han96}.

The study of Giant H~II regions (GH~II) in the near--infrared can
address important astrophysical questions such as: 1. characterizing
the stellar content by deriving the initial mass function (IMF), star
formation rate and age; 2. determining the physical processes involved
in the formation of massive stars, through the identification of OB
stars in very early evolutionary stages, such as embedded young
stellar objects (YSOs) and ultra--compact H~II regions (UCH~II); and
3. tracing the spiral arms of the Galaxy by measuring spectroscopic
parallaxes of zero age main sequence OB stars.  The exploration of the
stellar content of obscured Galactic GH~II regions has been studied
recently by several groups: \citet{han97}, \citep[2000, 2001]{blum99}, 
\citet{okum00}. In particular, Blum and collaborators presented near 
infrared imaging and spectroscopic observations of three optically 
obscured GH~II regions: W43, W42 and W31.  These observations revealed 
massive star clusters at the center of the H~II regions which had been 
previously discovered and studied only at longer wavelengths.

In this work, we present results for NGC3576 (G291.3-0.71), located at
a kinematic distance $2.8 \pm 0.3$ kpc, which we adopted from 
\citet{pre99}, after correcting for the standard Galactic center distance
($R_0 = 8$ kpc).  NGC3576 appears in the visible passbands as a faint
H~II region, but in the infrared it is among the most luminous in our
Galaxy \citep{gos69}.  In fact, with 1.6 x 10$^{50}$photons s$^{-1}$
inferred from the radio data, it can be classified as a GH~II
(following the suggestion of R. Kennicutt for sources brighter than
$10^{50}$ Lyman continuum, $=$ $L_{yc}$, photons per second, private
communication). A GH~II region has at least ten times the luminosity 
of the Orion nebula and
roughly the number emitted from the hottest single O3-type star, thus
implying multiple hot stars.

NGC3576 was observed in the radio continuum by \citet{gos70}.
\citet{gee68}, \citet{gee81}, \citet{wil70}, and \citet{pre99} have
detected radio recombination lines. Maser sources have also been
detected in the region: CH$_{3}$OH \citep{cas95} and H$_{2}$O
\citep{cas89}. The detection of intense emission in the 10 $\mu$m
window, H$_{2}$O masers, and the compact thermal emission in the radio
are typical indications of the primitive stages of star formation and
of a dense circumstellar environment.  Photometry from 1 to 2.5 $\mu$m
of the brightest sources was performed by Moorwood \& Salinari (1981)
and by \citet{per94} showing that the spectral energy
distributions of these objects suggests that they have excess
emission.  An intense CO J=2--1 line at 230 GHz was observed by
\citet{whi83} in the core region of NGC3576.

In the present paper, we present an investigation of the stellar
content of NGC3576 through the $J$, $H$ and $K$ imaging and $K-$band
spectroscopy (described in \S2).  In \S3 we discuss our results, and
our conclusions are summarized in \S4.

\section{Observations and Data Reduction}

$J$ ($\lambda$ $\approx$ 1.3 $\mu$m, $\Delta$$\lambda$ $\approx$ 0.3
$\mu$m), $H$ ($\lambda$ $\approx$ 1.6 $\mu$m, $\Delta$$\lambda$
$\approx$ 0.3 $\mu$m) and $K$ ($\lambda$ $\approx$ 2.1 $\mu$m,
$\Delta$$\lambda$ $\approx$ 0.4 $\mu$m) images of NGC3576 were
obtained on the nights of 1999 March 3 and 4 and 2000 May 19 and 20
with the f/14 tip--tilt system on the Cerro Tololo Interamerican
Observatory (CTIO) 4-m Blanco Telescope using the facility imager
OSIRIS\footnote{OSIRIS is a collaborative project between Ohio State
University and CTIO. Osiris was developed through NSF grants AST
90-16112 and AST 92-18449.})  and on the nights of 1998 July 9 to 13
with the facility imager CIRIM\footnote{CIRIM and OSIRIS are described
in the instrument manuals found on the CTIO WWW site at
www.ctio.noao.edu. For OSIRIS, see also \citep{poy93}.})  mounted on
the 1.5-m telescope. Spectroscopic data were obtained with the Blanco
telescope and the facility near infrared spectrometer, IRS, in 1998
May 17 and June 2--3, and with OSIRIS using the f/14 tip-tilt system
in 1999 March 3--4, 1999 May 2--3, and 2001 July 7 and 12.  OSIRIS
delivers a plate scale of 0.16$''$/pixel, the IRS 0.32$''$/pixel, and
CIRIM 1.16$''$/pixel.

All basic data reduction was accomplished using IRAF\footnote{IRAF is
distributed by the National Optical Astronomy Observatories.}. Each
image was flat-fielded using dome flats and then sky subtracted using
a median-combined image of five to six frames. Independent sky frames
were obtained 5--10$'$ south of the NGC3576 cluster.

\subsection{Imaging} 

The OSIRIS 1999 March images were obtained under photometric
conditions.  Total exposure times were 180s, 45s and 45s at $J$, $H$
and $K$, respectively. The individual $J$, $H$ and $K$ frames were
shifted and combined. These combined frames have point sources with
FWHM of $\approx$ 0.61$"$, 0.88$"$ and 0.64$"$ at $J$, $H$ and $K$,
respectively. DoPHOT \citep{sch93} photometry was performed on the
combined images. The flux calibration was accomplished using standard
star GSPC~S427-D (also known as [PMK98] 9123) from \citet{per98} which
is on the Las Campanas Observatory photometric system (LCO). The LCO
standards are essentially on the CIT/CTIO photometric system
\citep{eli82}, though color transformations exist between the two
systems for redder stars. No transformation exists between OSIRIS and
either CIT/CTIO or LCO systems.

The standard observations were made just after the NGC3576 data and
within 0.24 airmass from the target. No corrections were applied for
these small difference in airmass. Aperture corrections measured
inside 20 pixel radius circles were used to put the instrumental
magnitudes on a flux scale.  Six uncrowded stars on the NGC3576 images
were used for this purpose.  Since the brightest stars in the 1999
images were saturated, we have taken short exposure images in May
2000.  Although the conditions in May 2000 were non-photometric, we
used stars in common with the 1999 images to determine the zero point
for the additional (bright) stars.

Uncertainties for the $J$, $H$ and $K$ magnitudes in 1999 images
include the formal DoPHOT error added in quadrature to the error in
the mean of the photometric standard and to the uncertainty of the
aperture correction used in transforming from the DoPHOT photometry to
OSIRIS magnitudes. The sum in quadrature of the aperture correction
and standard star uncertainties are $\pm$0.032, $\pm$0.034 and
$\pm$0.069 mag in $J$, $H$ and $K$, respectively. The scatter in the
instrumental magnitudes in the set of stars from the 1999 images used
to calibrate the May 2000 images are $\pm$0.010 ($J$), $\pm$0.012
($H$) and $\pm$0.010 ($K$) mag; thus the errors in the bright star
magnitudes are dominated by the uncertainty in the standard stars.
The DoPHOT errors were larger than $\pm$ 0.01 mag, and we adopted a
cutoff for errors larger than 0.05 mag, which corresponds to a
limiting magnitude of $J,H$ $\sim$ 15.0.

Lower angular resolution images were obtained at $J$, $H$ and $K$
using CIRIM at f/8 on the CTIO 1.5-m telescope(1.16$"$
pixel$^{-1}$). The individual frames in each filter were shifted and
combined and have measured seeing of 2.2$"$ FWHM. Although collected
under photometric conditions, these images are not as deep as that of
the 4-m telescope and were used only to transform to equatorial
coordinates, since they encompassed a wider field than the OSIRIS
images taken at the 4-m telescope.

\subsection{Spectroscopy} 

The $K-$band spectra of eight of brightest stars in the NGC3576
cluster were obtained: \#11, \#48, \#69 and \#160 with the IRS and
\#4, \#48, \#78, \#95 and \#184 with OSIRIS. The spectra were divided
by the average continuum of sevaral A--type stars to remove telluric
absorption.  The Br$\gamma$ photospheric feature was removed from the
average A--type star spectrum by eye by drawing a line between two
continuum points.  One dimensional spectra were obtained by extracting
and summing the flux in $\pm$ 2 pixel aperture. The extractions
include background subtraction from apertures, 1-2$''$ on either side
of the object.

The wavelength calibration was accomplished by measuring the position
of bright OH$^{-}$ lines from the $K-$band sky spectrum \citep{oli92}.
The spectral resolution at 2.2 $\mu$m is $\lambda$/$\Delta$$\lambda$
$\approx$ 3000 for OSIRIS and $\lambda$/$\Delta$$\lambda$ $\approx$
825 for the IRS.

\section{Results}

\subsection{Near-infrared Imaging}

The OSIRIS $J$, $H$ and $K-$band images reveal an embedded star
cluster.  We detected 315 stars in the $K-$band to a limiting
magnitude of 16 (see below) in a region of 0.03 square degrees.
Figure~\ref{finding} shows a finding chart using the $K-$band image. A
false color image is presented in Figure~\ref{color}, made by
combining the three near infrared images and adopting the colors blue,
green and red, for $J$, $H$, and $K$, respectively. In this way, the
bluest stars are likely foreground objects, and the reddest stars are
probably $K-$band excess objects, indicating the presence of hot dust
for objects recently formed in the cluster (background objects seen
through a high column of interstellar dust would also appear red). The
diffuse nebula is mainly due to Br$\gamma$ emission in the H~II
region. The dark patches in the bottom right of Figure~\ref{color} are
zones of the giant molecular cloud from which NGC3576 is
emerging. There is no doubt that this is a signature of a young
cluster containing massive stars, now in the process of shredding the
local molecular cloud. The cluster is asymmetric, with the majority of
the stars encompassed in a semi circle; and there is a definite
appearance that the cluster is destroying the cloud from the NE to the
SW. The NE is also the direction toward which the H~II region is seen
in the visual.

The $H - K$ {\it{versus}} $K$ color magnitude diagram (CMD) is
displayed in Figure~\ref{cmd}. The open circles indicate objects
fainter than $H = 15$.  A concentration of dots appear around $H - K =
1$, probably indicating the average color of cluster members. A number
of stars display much redder colors, especially the brightest ones.
The solid vertical line indicates the theoretical main sequence (see
below).

\begin{figure}[ccd]
\plotone{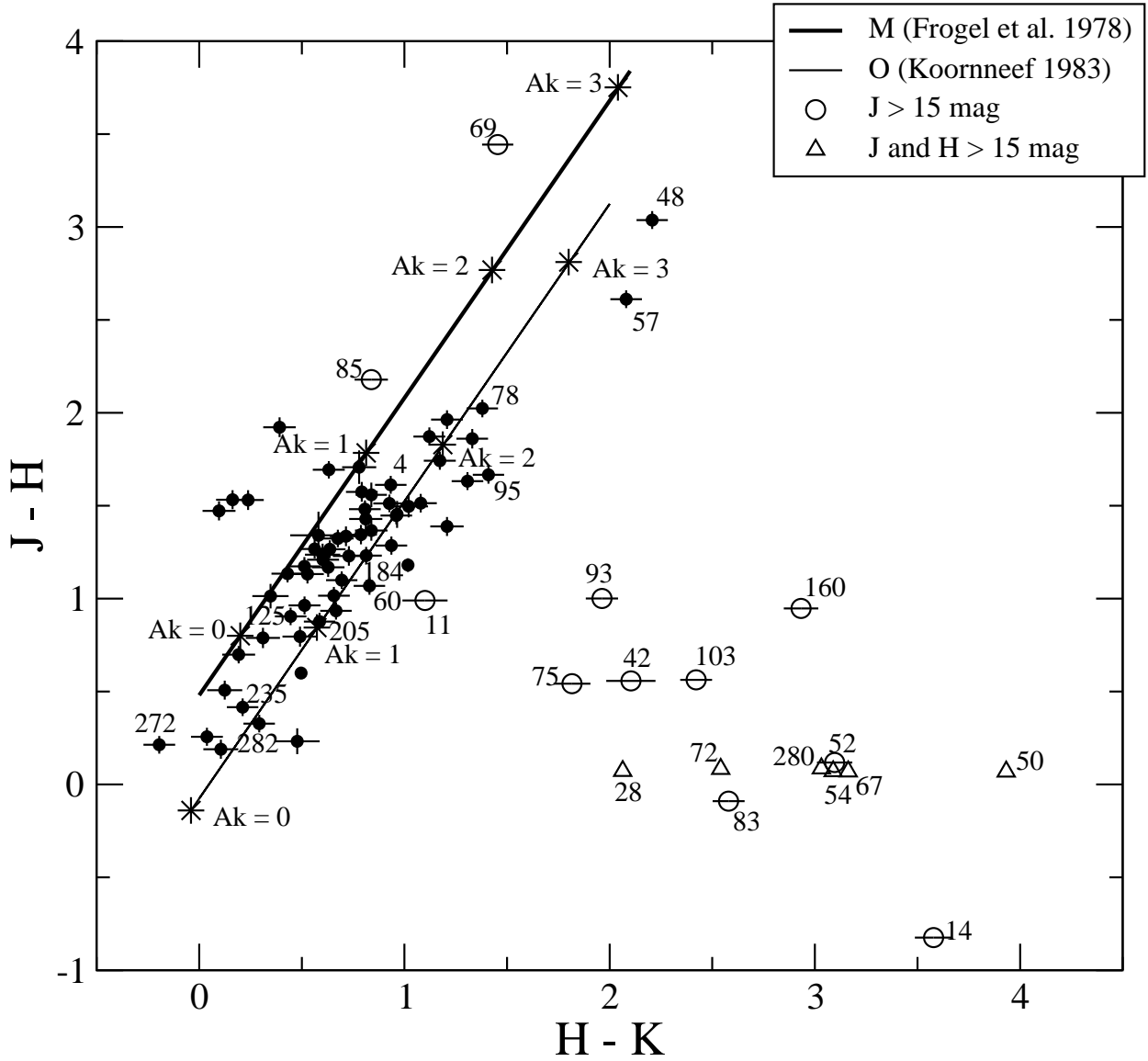}
\caption{$J - H$ vs $H - K$ color--color plot showing the reddening
line of M--type stars ({\it heavy solid} line) and O--type stars ({\it
solid} line). {\it Open triangles} indicate stars with $J$ and $H > 15$; 
{\it open circles}, stars with $J > 15$. The {\it asterisks} refer to $A_{K}$
reddening. Object labels are the same as in Figure~1. \label{ccd}}
\end{figure}

The $J - H$ {\it{versus}} $H - K$ color--color plot is displayed in
Figure~\ref{ccd}. Open circles indicate objects fainter than $J = 15$.
Open triangles indicate stars fainter than $H = 15$ and $J = 15$. The
numbers labeling stars in both plots refer to the same objects. The
inclined lines, from top to bottom, indicate interstellar reddening
directions for main sequence M--type \citep{fro78}, O--type
\citep{koo83} and T~Tauri \citep{mey97} stars.  Stars to the right of
the solid line deviate from pure interstellar reddening, probably
because of hot dust emission.  The effect of this excess emission is
stronger in the $K-$band than at shorter wavelengths. Although the
open circles and triangles in the bottom right of Figure~\ref{ccd}
indicate only lower limits, these objects are also likely affected by
thermal emission.  They are bright in the $K-$band and should be
detected in $J$ and/or $H$ if affected only by interstellar reddening
and if they were cluster members with typical extinction.

\subsubsection{Reddening and excess emission } 
 
The cluster characteristics indicate that the stars haven't had enough
time to evolve away from the main sequence. We expect that most of the
high mass stars are close to the Zero Age Main Sequence (ZAMS).  We
can estimate the reddening toward the cluster from a simple
approximation \citep{mat90} $A_K$ $\sim$ 1.6$\times (H - K)$ and
using the fact that the average intrinsic color of hot stars is almost
zero \citep{koo83}. The stars brighter than $K = 14$ have an average
color of $H - K = 0.98$, corresponding to $A_{K}$ = 1.57 mag ($A_{V}$
$\approx$ 15.7 mag). The interstellar component of the reddening can
be separated from that local to the cluster stars by using the star HD
97499. This is a foreground star, since it is brighter and less
reddened than stars in the cluster and is offset from the radio source
line-of-sight by 2 arc-minutes. From its spectral type B1-2IV-V
{\it{Michigan Spectral Catalogue}} - \citep{hou75} and magnitude, a
distance of 2.4 kpc was derived \citep{per94}. Our measurements of
this star result in $H - K = 0.19$, indicating A$_{K} = 0.37$. This
gives A$_{K} = 0.154$mag kpc$^{-1}$, which is close to the expected
extinction for this position along the Galactic plane. For 2.8 kpc
distance of NGC3576, the interstellar component is then A$_{K} =
0.43$, leaving a local component of A$_{K} = 1.14$. We were not able
to independently check the distance derived by DePree (1999) via
spectroscopic parallax (see below), but we expect that the radio
kinematic distance is reliable for this Galactic direction.

In order to place the ZAMS in the CMD, the corresponding bolometric
magnitudes ($M_{\rm Bol}$) and effective temperatures ($T_{eff}$) must
be transformed into $H - K$ colors and apparent $K$
magnitudes. \citet{sch92} models for the ZAMS with solar abundances
were adopted. The bolometric corrections (BC) applied to derive
absolute visual magnitudes ($M_V$) are from \citet{vac96} for spectral
types $O$ to $B0$ and from \citet{mal86} for later spectral types. The
intrinsic colors ($(V - K)_{0}$, $(J - K)_{0}$, and $(H - K)_{0}$) are from
\citet{koo83}. Using the distance modulus (12.24) and the apparent
visual magnitude, we transform the $M_{\rm Bol}$ into apparent $K_{0}$
magnitudes. The correspondence between spectral types and $T_{eff}$
are from \citet{vac96} for $O$ to $B0$ and from \citet{joh66} for
later spectral types. Since Koornneef's colors are in the Johnson
system, which is nearly identical to SAAO system \citep{car90}, we
used the SAAO to CIT/CTIO relations to transform Koornneef's $H - K$
color indices to the CIT/CTIO system. These corrections are about 1\%
and could be neglected when compared with the photometric errors and
differential reddening.

The ZAMS is represented by a vertical solid line in Figure~\ref{cmd},
shifted to $D = 2.8$ kpc and reddened by A$_{K}$ = 0.43 due to the
interstellar component.  When adding the average local reddening
(A$_{K} = 1.14)$, the ZAMS line is displaced to the right and down, as
indicated by the dashed lines.  We cannot fix the position of the ZAMS,
since there is a scatter in the reddening. The small group of
relatively bright stars ($K \sim 11$) in between these two lines,
suggests that some of them, the bluer ones, could mark the position of
the ZAMS. Unfortunately, we don't yet have spectra of these objects to
check whether or not this is true.

Objects to the right of the O-type stars line (Figure~\ref{ccd}) have
colors deviating from pure interstellar reddening. This is frequently
seen in young star clusters and is explained by hot dust in the
circumstellar environment. We can estimate a lower limit to the excess
emission in the $K-$band by supposing that the excess at $J$ and $H$
are negligible, and that the intrinsic colors of the embedded stars
are that of OB stars. Indeed, assuming that our sample of stars is
composed by young objects (not contaminated by foreground or
background stars), any object would have an intrinsic color in the
range $(H - K)_{0} = 0.0 \pm 0.06$ mag \citep{koo83}.  Let us adopt
for all objects in our sample the intrinsic colors of a B2 V star: $(J
- H)_{0} = -0.09$ and $(H - K)_{0} = -0.04$ \citep{koo83}. The error
in the color index would be smaller than the uncertainty in the Mathis
law, we are using for the interstellar extinction.  >From the
difference between the observed $J - H$ and the adopted B 2 V star we
obtain the color excess and by using the relation ($A_{J} = 2.58
\times E_{J-H}$) we derive the interstellar extinction in the $J-$band
and so the intrinsic apparent magnitudes $J_{0}$, $H_{0}$ and $K_{0}$.
>>From another relation of Mathis law ($A_{K} = 0.382 \times A_{J}$),
we derive the magnitude at $K-$band corrected from the interstellar
extinction. The difference between this number and $K_{0}$ gives the
excess emission in the $K-$band, due to dust thermal emission.

\begin{figure}[Kexc]
\plotone{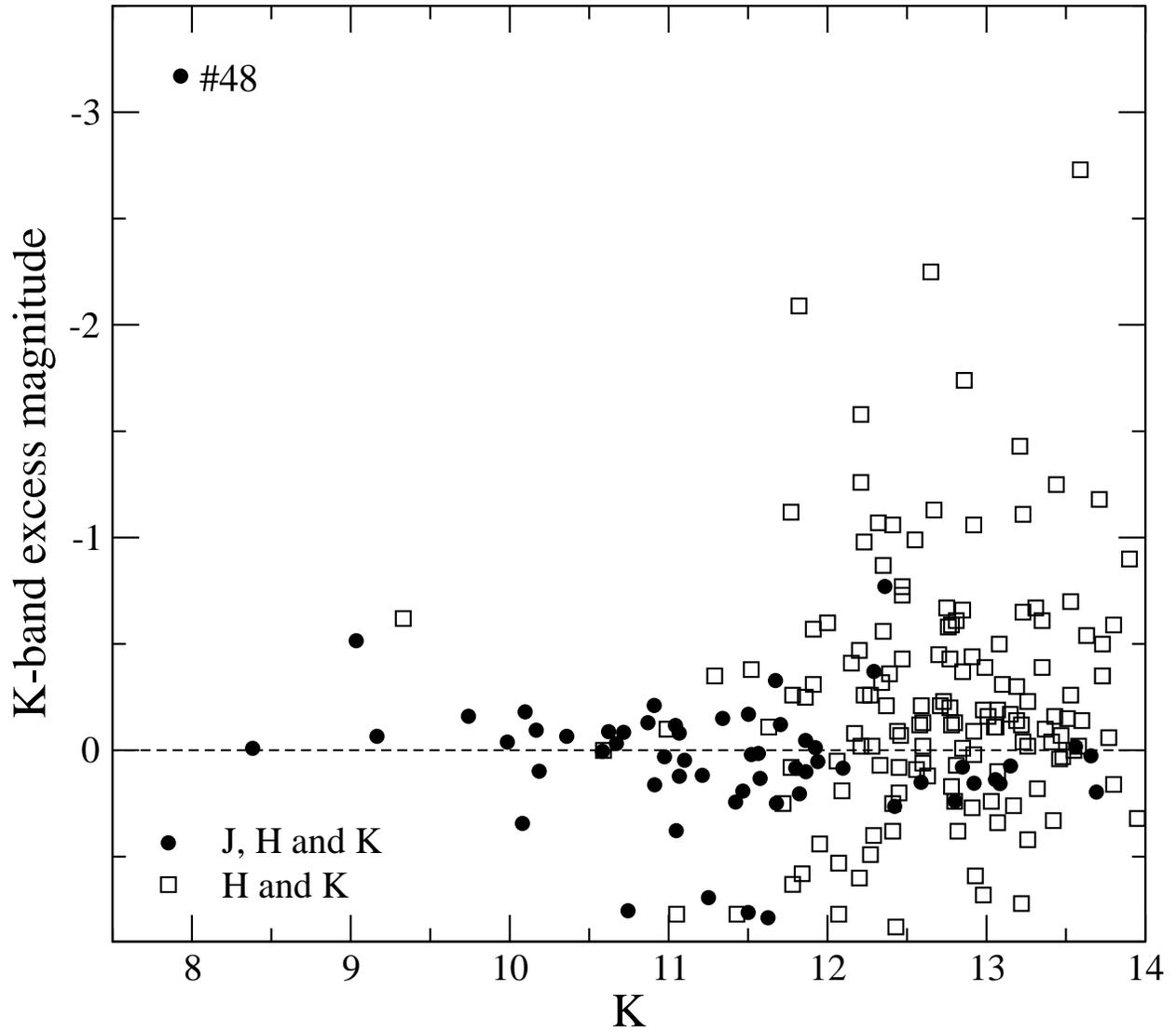}
\caption{Excess emission as a function of derredned $K-$band magnitude
($K_{\circ}$). 
Very negative values represent circumstellar emission. \label{Kexc}}
\end{figure}

For stars not detected in the $J-$band we suppose they are affected by
an interstellar extinction $A_{J}$ equal to the median value of those
measured in the $J-$band. In this way, we derive the extinction in the
$H-$band by using Mathis law in the form: $A_{H} = 0.624 \times
A_{Jmedian}$. From here the procedure to derive the $K-$excess
magnitude follows the same steps as we did before.  For stars not
detected in the $H-$band, we still want to estimate the $K-$excess,
since potentially interesting objects are too red to be detected in
the $H-$band.  A lower limit for this excess can be derived by using
the same procedure as above, but assigning a limiting magnitude $H =
15$ for objects not detected in the $H-$band. Our results are
displayed in Figure~\ref{Kexc}. Objects with very large excess in the
upper right corner of Figure~\ref{Kexc}, cannot be explained by errors
in the de--reddening procedure and could be real. They could represent
the emission of accreting disks around the less massive objects of the
cluster.

\subsubsection{The KLF and the IMF} 

In order to separate the cluster members from projected stars in the
cluster direction, we imaged a region close to NGC3576.  The star
counts were normalized by the relative areas projected on the sky and
then binned in intervals of $\Delta K = 0.5$ and $\Delta (H - K) =
0.5$. The stellar density in the field was then subtracted from that
of the cluster in bins of magnitude and color intervals.  This works
well for foreground stars, since there were a few in the cluster
field. Regarding the background, the situation is more complex,
however, since NGC3576 produces so high an obscuration, almost no
background objects can be seen at this limiting magnitude.
 
\begin{figure}[completeness]
\plotone{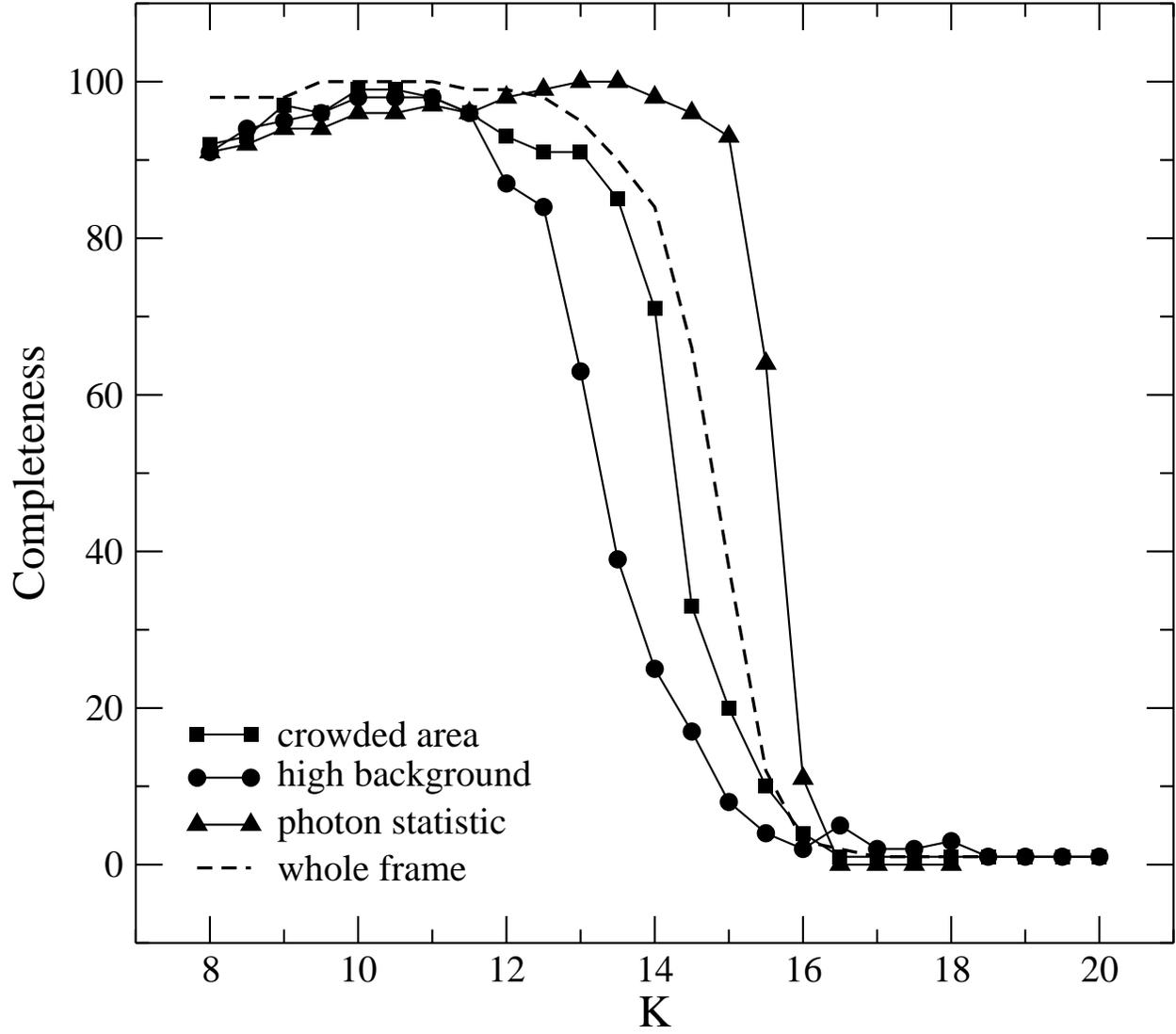}
\caption{Completeness (in percent detection) as derived from
artificial star experiment. The {\it dashed} line is for the whole
frame and the symbols are for representative sub--regions of the
frame; see text.
\label{completeness}}
\end{figure}

The completeness of DoPHOT detections was determined through
artificial star experiments. This was accomplished by inserting fake
stars in random positions of the original frame, and then checking how
many times DoPHOT retrieved them. The PSF of the fake star was
determined from an average of real stars found in isolation and in
areas of dark sky. In total, we inserted 2400 stars in the magnitude
interval 8 $\leq$ K $\leq$ 20, which amounts to six times the number
of real stars recovered in the original DoPHOT run. For every $\Delta
K = 0.5$ we inserted simultaneously 5 stars, repeating the procedure
for 20 times.  The insertion of all the 100 stars at once would impact
the stellar crowding and change the detection conditions. The
incompleteness of the sample is defined as the percentage of times the
fake star fails to be recovered. We performed these experiments in the
whole frame and also in each of the three sub--images that were cut
out from it, displayed in Figure~\ref{finding}. The upper right
sub--image is representative of detection limited by photon
statistics, the one at the center, by high background and the lower
left one, by stellar crowding.  In Figure~\ref{completeness} we
present the photometric completeness. The {\it dashed} line (without
symbols) refers to the whole image. The limit is different for
different sub--images, e.g. for an area with high nebular background
({\it circles}), for a crowded area ({\it squares}), or for an area
with few stars and a dark sky ({\it triangles}). The performance of
the photometry is better than 90\% for a 15th magnitude star found in
isolation, as compared to stars in the nebular zone which need to be
ten times brighter to be detected with the same efficiency. Since
there is no objective way to define the sub-image limits we applied a
single completeness correction to the whole frame (dashed line). As
seen in Figure~\ref{completeness} such a correction is close to the
curve limited by crowding. Future work seeking to obtain deeper
photometry in this cluster demands a substantial improvement of the
spatial resolution and will require adaptative optics imaging from
ground based telescopes.
 
\begin{figure}[klf]
\plotone{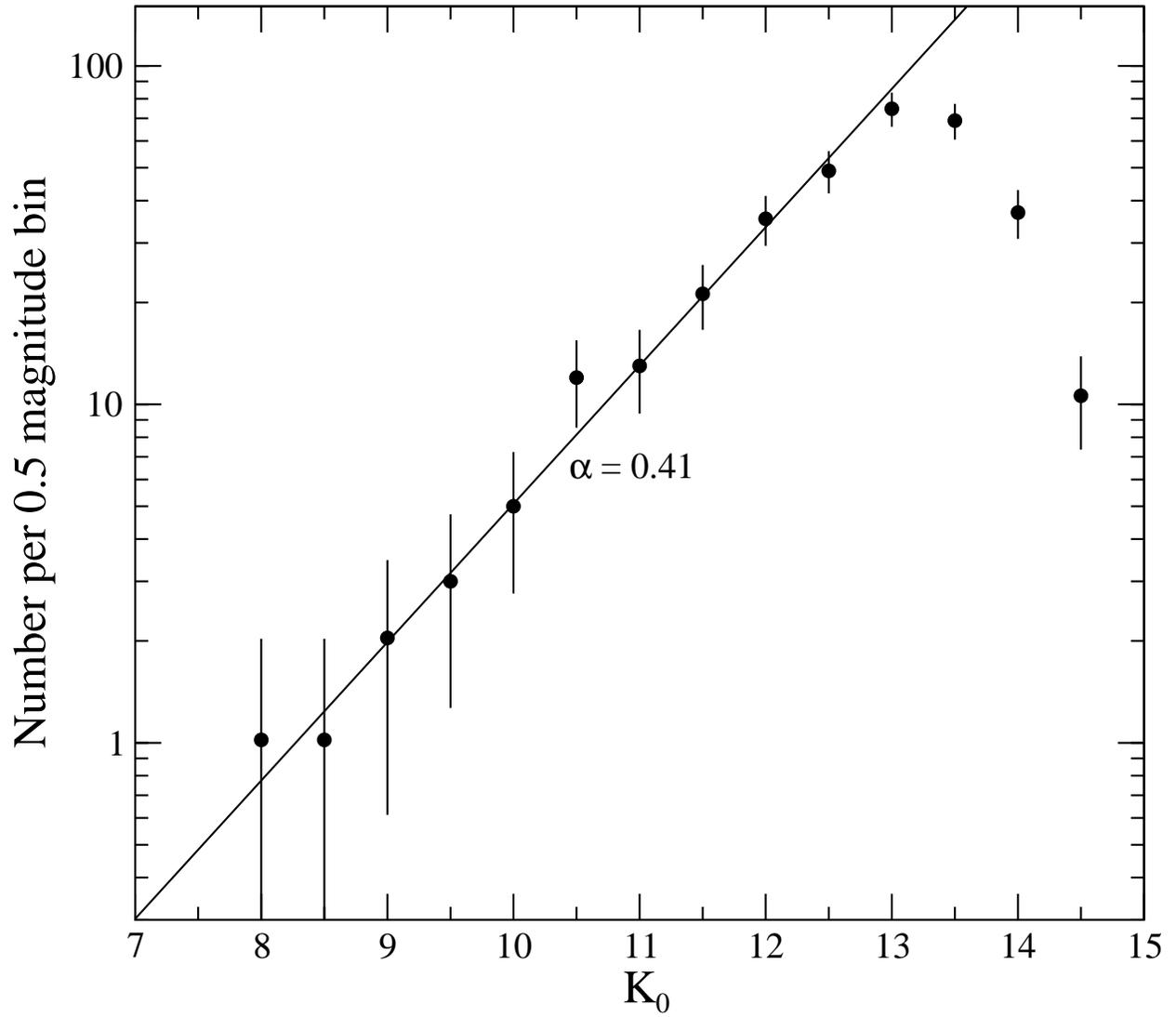}
\caption{The $K-$band luminosity function corrected for non-members
and sample incompleteness. $K_{\circ}$ is dereddened and corrected for
excess emission; see text. \label{klf}}
\end{figure}

After correcting for non-cluster members, interstellar reddening,
excess emission (a lower limit) and photometric completeness, the
resulting $K-$band luminosity function ($KLF$) is presented in
Figure~\ref{klf}.  A linear fit, excluding deviant measures by more
than $3 \sigma$, has a slope $\alpha = 0.41 \pm 0.02$. A similar KLF
slope was obtained for W42 ($\alpha = 0.40$) by \citet{blum00}.

We can evaluate the stellar masses by using \citet{sch92} models,
assuming that the stars are on the ZAMS and not the pre-main sequence
(but see below the discussion in \S3.2).  This is a reasonable
approximation for massive members of such a young cluster.  The main
errors in the stellar masses are due to the effects of circumstellar
emission and stellar multiplicity. Our correction to the excess
emission is only a lower limit, since we assumed the excess was
primarily in the $K$ band. \citet{hsvk92} have computed disk
reprocessing models which show the excess in $J$ and $H$ can also be
large for disks which reprocess the central star radiation. In
general, we can expect the excess emission to result in an
overestimate of the mass of any given star and the cluster as a
whole. The slope of the mass function should be less effected.  It is
difficult to quantify the effect of binarity on the IMF. If a given
source is binary, for example, its combined mass would be larger than
inferred from the luminosity of a ``single'' star and its combined
ionizing flux would be smaller. The cluster total mass would be
underestimated, the number of massive stars and the ionizing flux
would be overestimated.  The derived IMF slope would be flatter than
the actual one.

\begin{figure}[imf]
\plotone{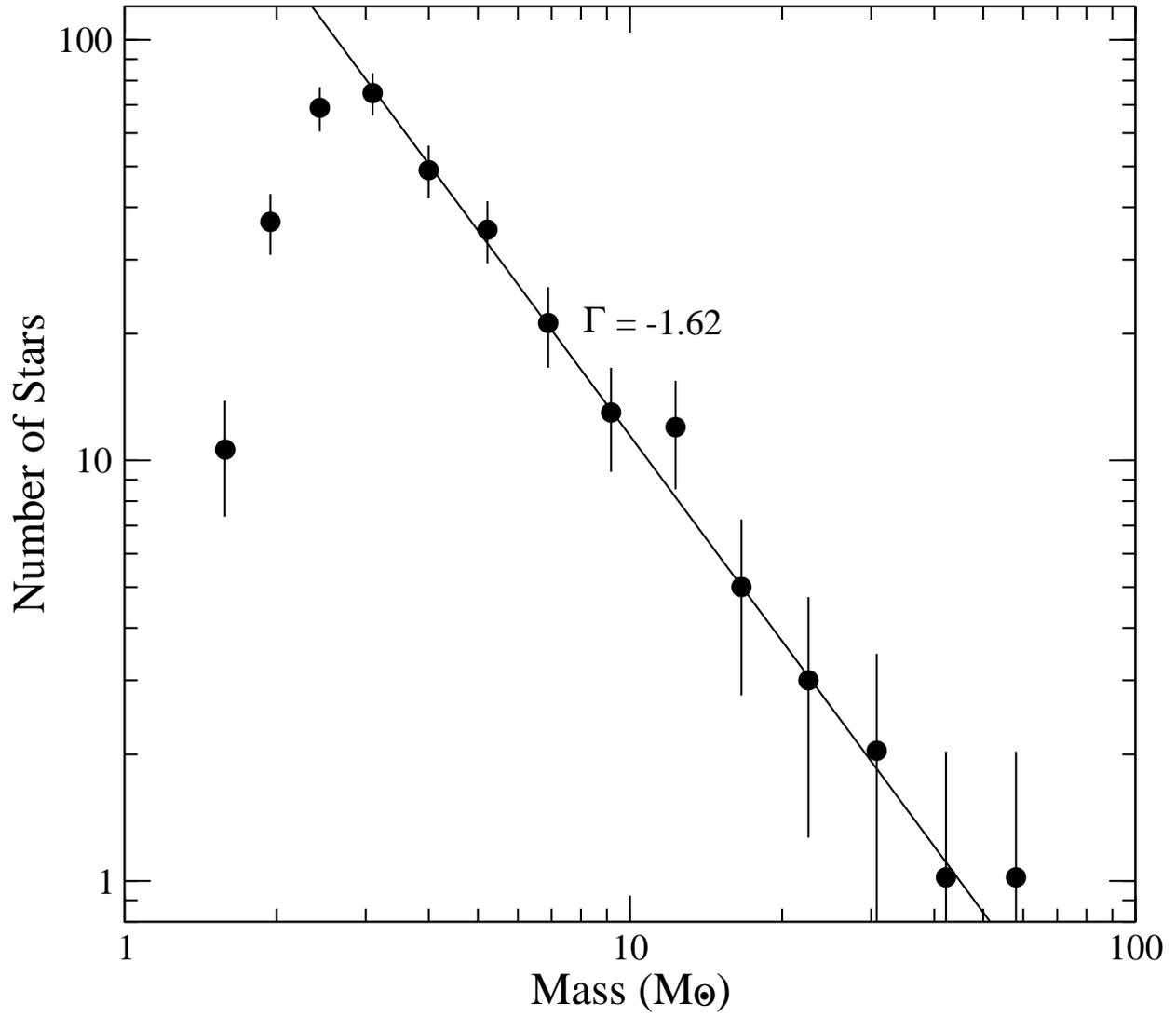}
\caption{The IMF of the cluster members using the models of
\citet{sch92} and the KLF from Figure~\ref{klf}. \label{imf}}
\end{figure}

With these limitations in mind we have transformed the KLF into an
IMF.  Since other authors also do not typically correct for
multiplicity, our results can be inter compared, as long as this
parameter doesn't change from cluster to cluster. The $IMF$ slope
derived for NGC3576 is $\Gamma = -1.62$ (Figure~\ref{imf}), which is
consistent with Salpeter's slope \citep{sal55}.  A similar IMF slope
was obtained for the Trapezium cluster ($\Gamma = -1.43$) by
\citet{hil00}.  Flatter slopes have been reported only for a few
clusters, most notably the Arches and Quintuplet clusters
\citep{fi99a}, both near the Galactic Center.  Flatter slopes may
indicate that in the inner Galaxy star forming regions, the relative
number of high mass to the low mass stars is higher than elsewhere in
Galaxy. It is also possible that dynamical effects may be more
important in the inner Galaxy. \citet{pz01} have modeled the Arches
cluster data with a normal IMF, but include the effects of dynamical
evolution in the presence of the Galactic center gravitational
potential. They find the observed counts are consistent with an intial
Salpeter--like IMF.

We derived an upper limit to the total mass of the cluster (our IMF is
likely overestimated due to excess emission, see above) by integrating
the $IMF$ between $0.08 < M/M_\odot < 58 $ - where the distribution is
nearly continuous. The lower mass limit was adopted from \citet{hil98}
taking into account the $IMF$ turnover measured by those authors in
Orion.  The integrated cluster mass is M$_{cluster}$ = 5.4 x 10$^3$
M$_\odot$. As pointed out above, this is likely an upper limit.

\subsubsection{The Nature of NGC3576/\#48}

Source \#48 is anomalously bright and needs to be treated
separately. Let us assume that it is affected by an interstellar
reddening equal to the cluster average: A$_{K} = 1.57$, which implies
a derredned magnitude K$_0 = 6.78$. If we take at face value the $H -
K$ color to represent only an excess at $K$, the reprocessing disk
would contribute with $\Delta K =- 0.3$. This excess emission is
clearly an underestimation, since the stellar flux is swamped by the
disk emission to the point of veiling all the photospheric lines (see
below).  Moreover, if the excess emission was so small, the luminosity
of \#48 would require a cluster of four 100 M$_\odot$ stars,
unresolved down to a limit of 0.6$''$, which doesn't seem to be the
case. If object \#48 is a single O3 V star, it would contribute with
NLyc $=$ 1.17 $\pm$ 0.05 x 10$^{50}$ s$^{-1}$.
 
Alternately, we can evaluate the excess emission by using
\citet{hsvk92} models for reprocessing disks.  By starting with the
maximum excess emission $\Delta K = 4.05$ valid for a O7-type star
(their table 4) we derive the M$_V$.  Using Vacca et al. (1996)
calibration, we obtain the corresponding stellar spectral type (and
mass) that is much smaller than the O7-type we started with. The next
step is reducing the excess emission (adequate for a smaller stellar
luminosity), deriving a larger final mass, and iterating until
convergence.  This was achieved for an excess emission $\Delta K =
-3.17$, corresponding to a spectral type B1 V and mass M $=$ 17
M$_\odot$. This is possibly a lower limit, since the intervening
extinction toward \#48 is probably larger than the cluster average. We
can say that \#48 is a late--O/early--B/early B YSO, very similar to
what has been found by \citet{blum00} in W31 for the brightest
$K-$band object in that cluster. Object \#48 is buried in a large and
dense disk, and its contribution to the cluster mass and ionizing
photons are negligible.

We have compared the locations of the $K-$band sources in our images
with the mid--infrared sources of \citep{per94}. Source \#48 is very
close to their IRS1 and \#11 to their IRS3. These sources appear also
in the IRAS Small Scale Structure Catalogue (X1109-610) and the IRAS
Point Source Catalogue (IRAS11097-6102).  In the case of IRS1, at
least, recent high resolution mid-infrared images \citep{B2002}
clearly indicate that it is associated with our object \#50, 1$''$ to
the south of the $K-$band source \#48.

\subsubsection{General Properties of the Stellar Cluster}

The number of Lyman continuum photons derived from the IMF, excluding
object \#48 is NLyc $=$ 0.42 $\pm$ 0.22 x 10$^{50}$ s$^{-1}$.  The
contribution of this single object could be as large as NLyc $=$ 1.17
x 10$^{50}$ s$^{-1}$, in the case it is an O3 V star. This is very
close to the NLyc $=$ 1.6 $\pm$ 0.4 x 10$^{50}$ s$^{-1}$ derived from
radio observations \citep{pre99} scaled to a distance of 8 Kpc to the
Galactic center. However, we have shown in the preceeding sub-section,
object \#48 probably is a much less massive object, a B1 V star.  It
can be seen that the NLyc is highly sensitive to the particular
procedure used to correct for the excess emission and extinction. Very
probably we have missed a handful of main sequence O-type stars
responsible for the ionizing flux seen at radio wavelengths. Moreover,
the spectra of the eight brightest stars described in the next
section, indicate that the ionizing stars must be apparently faint.
The properties of the cluster are summarized in Table~1.

\begin{figure}[cc_map]
\plotone{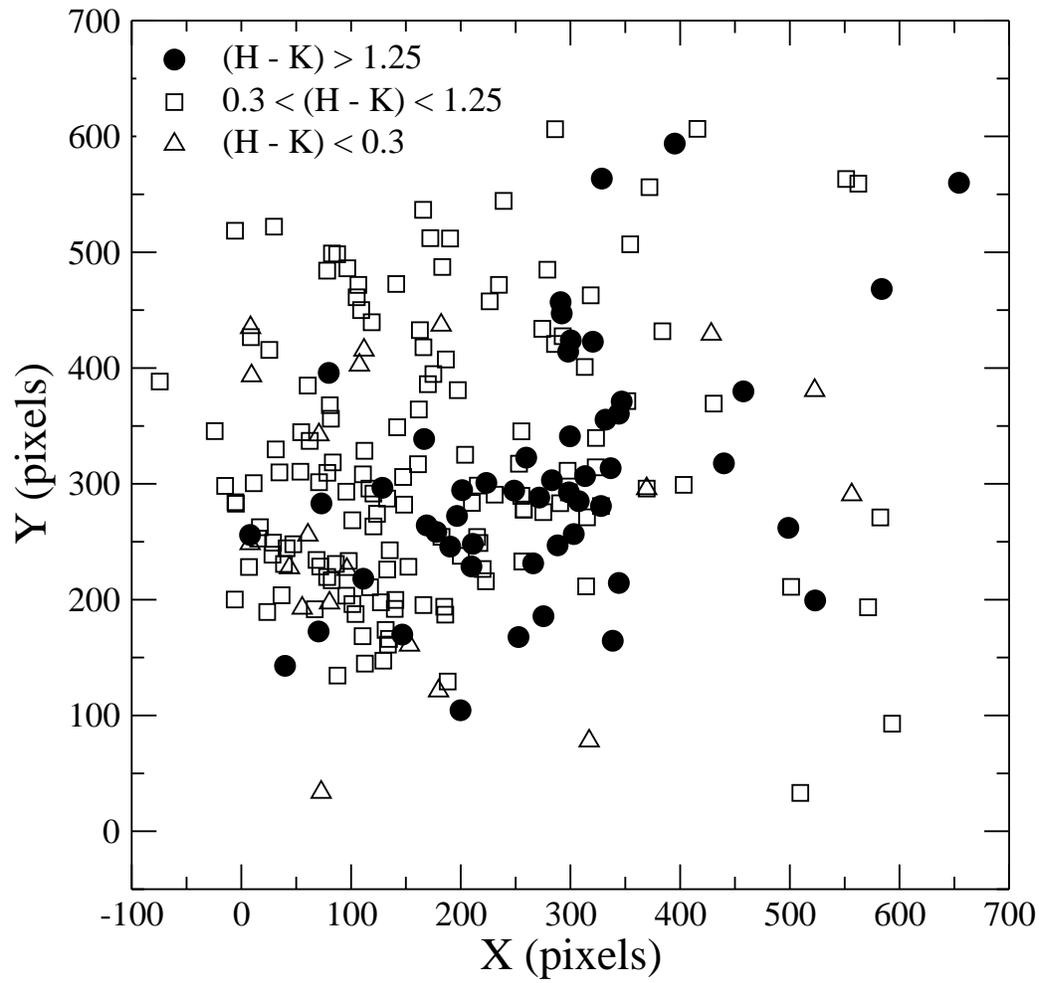}
\caption{Spatial distribution of the $H - K$ color index. North is 
up and East to the left. An increase in the color excess is seen 
toward SW (bottom right).  \label{cc_map}}
\end{figure}

The stellar cluster is located in the NE border of a molecular
cloud. The stellar density increases toward the SW, ending abruptly,
with a few sources embedded in the molecular cloud
(Figure~\ref{color}). The spatial distribution of color indices
(Figure~\ref{cc_map}) also shows a similar gradient. This is due, in
part, to the increasing extinction toward the inner molecular
cloud. However, the red colors are intrinsic to many of the sources,
since there are excess emission objects in this zone. This suggests
that stars at the SW are younger than at NE, since recent models by
\citet{bm01} predict formation times of the same order for stars of
different masses. A similar scenario for this cluster, by which star
formation is progressing toward the inner zones of the molecular
cloud, has been suggested by \citet{per94} and \citet{pre99} and our
images dramatically confirm this to be the case; see
Figure~\ref{color}.
 
\subsection{Spectra} 

\begin{figure}[ftless]
\plotone{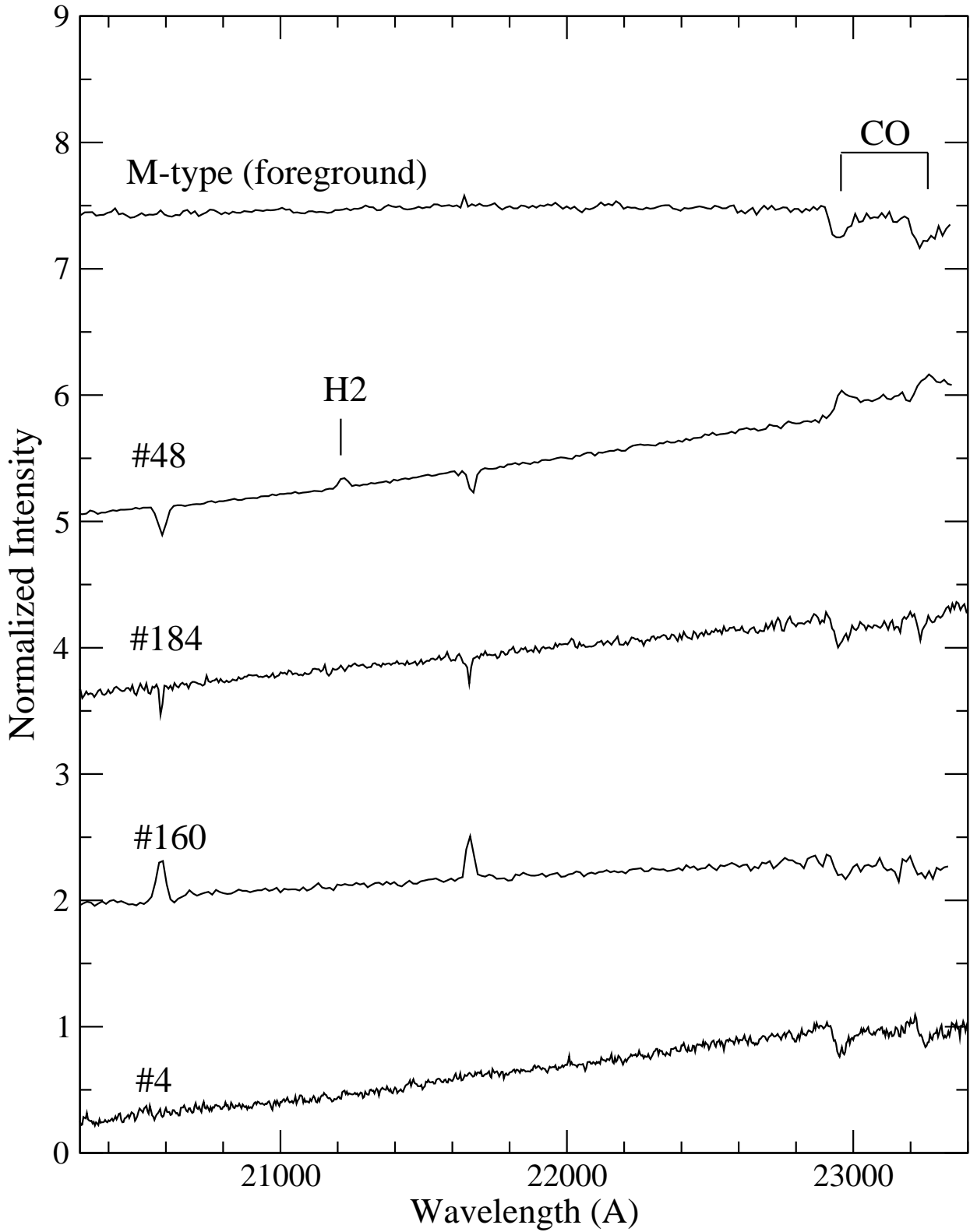}
\caption{Spectra of four cluster members displaying featureless
continua.  Objects \#11, \#69, \#78 show H$_2$ in emission. The
absorption feature in \#95 is due to over subtraction of the nebular
component.  Stars \#11 and \#69 were observed with the IRS and stars
\#78 and \#95 with OSIRIS. Fluxes were divided by the average 
continuum of several A--type stars and normalized to unity at 2.19 $\mu$m.  
The spectrum at the top is a foreground A3V-type star 
(also divided by the average A--type star continuum).\label{ftless}}
\end{figure}

Spectra of eight cluster members (\#4, \#11, \#48, \#69, \#78, \#95,
\#160 and \#184), are shown in Figure~\ref{co} and
Figure~\ref{ftless}.  Spectra of a foreground M-type and an A-type
stars were added at the top of the figures for comparison. The A-type
star also was divided by the average continuum slope of the other
observed A-type stars.

Object labels are the same as in Figure~\ref{finding};~\ref{cmd};
and~\ref{ccd}. Ordinates in Figure~\ref{co} and Figure~\ref{ftless}
are normalized fluxes, as follows.  We constructed templates for
telluric absorption bands by observing A-type stars, close in time and
airmass to the target stars. We removed the Br$\gamma$ line from those
spectra by hand (linear interpolation), since this region is free from
telluric features. There are no other noticeable photospheric features
in A-type stars in the $K-$band.  Then we divided the spectrum of each
target by the appropriate A-type spectrum and normalized the resulting
spectrum at 2.19 $\mu$m.  The signal--to--noise ratio is S/N $\approx
40$ for \#4, \#11 and \#95 and $>100$ for the other objects.
  
The spectra were placed on a flux scale by dividing by $\lambda^4$ and
multiplying by $K-$band fluxes corresponding to the $K-$magnitudes in
Table~2. However, it is straight forward to compare the observered
spectra ratioed only by the A--type continuum.  It can be seen that
all the cluster member candidates display rising continuua to the red
when compared to the A--type and M--type foreground stars. These eight
stars are thus most likely cluster members, but we can not rule out
that some may be background stars. We must take into account that
these objects are projected toward the central part of the cluster and
thus subject to high obscuration (A$_{K} = 1.57$) due to the intra
cluster gas and dust. Objects \#48, \#95, and \#160 are undoubtedly
cluster members, since, in addition, they have excess emission. Object
\#69 is bright in the $H$ and $K$ bands and judging from its $H - K$
color, it should be relatively bright in the $J-$band, but it isn't
detected at all in the $J-$band OSIRIS images, appearing above the
reddening line in Figure~\ref{ccd}. The spectrum of this object in
Figure~\ref{ftless} looks like its neighbor \#78, that is 1.5
magnitude fainter and still is detected in the $J-$band. This puzzling
situation was clarified when we took a $K-$band acquisition image
(Feb/2002) with Phoenix at Gemini under very good seeing ($\sim$
0.25--0.3''). In that image, object \#69 is shown as a small nebula,
with no sign of a buried point-like source. Object \#69 is a clump of
dense material ionized by object \#48 or some other neighboring
source.  The H$_2$ 2.122 $\mu$m emission could be either shock or
ionization produced.

The narrow features at 2.058 (\ion{He}{1}) and 2.166 $\mu$m
(Br$\gamma$) in Figures~\ref{co} and~\ref{ftless} are due to
contamination from the extended nebula.  The spectra were extracted in
such a way that the large scale nebular component is
over--subtracted. In some objects these lines are in emission, due to
enhanced nebular emission close to the star. Objects \#48, \#160, \#78
and \#69 show H$_2$ 2.122 $\mu$m in emission.  This feature appears is
emission also in \#11 and \#95, but since the He~I and Br$\gamma$
components are not over subtracted, this feature may be due to
contamination of extended nebular emission.  The CO bandhead at 2.2935
$\mu$m is in absorption in \#4, \#160, \#184 and in emission in \#48.
None of those objects show photospheric lines indicating that they are
still enshrouded in their birth cocoons. This is corroborated by the
excess emission in the $K-$band derived from photometry except for
objects \#4 and \#184 (Table~2).

A variety of mechanisms and models have been proposed to explain the
origin of CO emission in YSOs. These include circumstellar disks,
stellar or disk winds, magnetic accretion mechanisms such as funnel
flows, and inner disk instabilities similar to those which have been
observed in FU Orionis--like objects and T~Tauri stars in a phase of
disk accretion \citep{carr89,carr93,chan93,bis97}. \citet{mar97} shows
that gas free-falling along the field lines yield the bandhead
profiles, in agreement with those observed, with the shape of the
profile determined mainly by inclination of the disk to the line of
sight. Hanson et al. (1997) reported the presence of CO in emission in
several masssive stars in M17. The situation is less clear for objects
\#4, \#160 and \#184 displaying CO in absorption.  The absence of
large color excess indicates that they could be cool pre-main sequence
stars still in a contraction phase. While we can not rule out that
they are evolved background M-type stars, they do appear projected on
the core of the newly formed cluster. These objects deserve further
study, and if they are indeed pre-mainsequence stars, then the IMF
determination above has a component whose masses have been
overestimated.

The H$_2$ molecular emission is produced by shocks and may indicate
the existence of gas outflow. Since we subtracted the extended
background close to the stars, the spectra show only the spatially
unresolved component of H$_2$. We can thus be confident that H$_2$
emission in \#48 and \#69 are emitted close to the stellar
sources. However, since these two sources are only $\sim 11''$ apart,
the emission could be associated with either or both objects.

It may be surprising that most of the stars with featureless spectra
are close to the interstellar reddening line. However this is in
accord with the Hillenbrand et al. models (1992) for early type stars,
which predict relatively small color excesses ($\Delta ($H - K$) <
0.5$) for objects with large excess emission ($\Delta K < 4.0$).  Such
models do not predict large departure from the reddening line, like
displayed by objects \#160 and \#50. Those sources might be surrounded
by local dusty clouds, in addition to the accreting disk.

\section{Discussion}

\begin{figure}[color]
\plotone{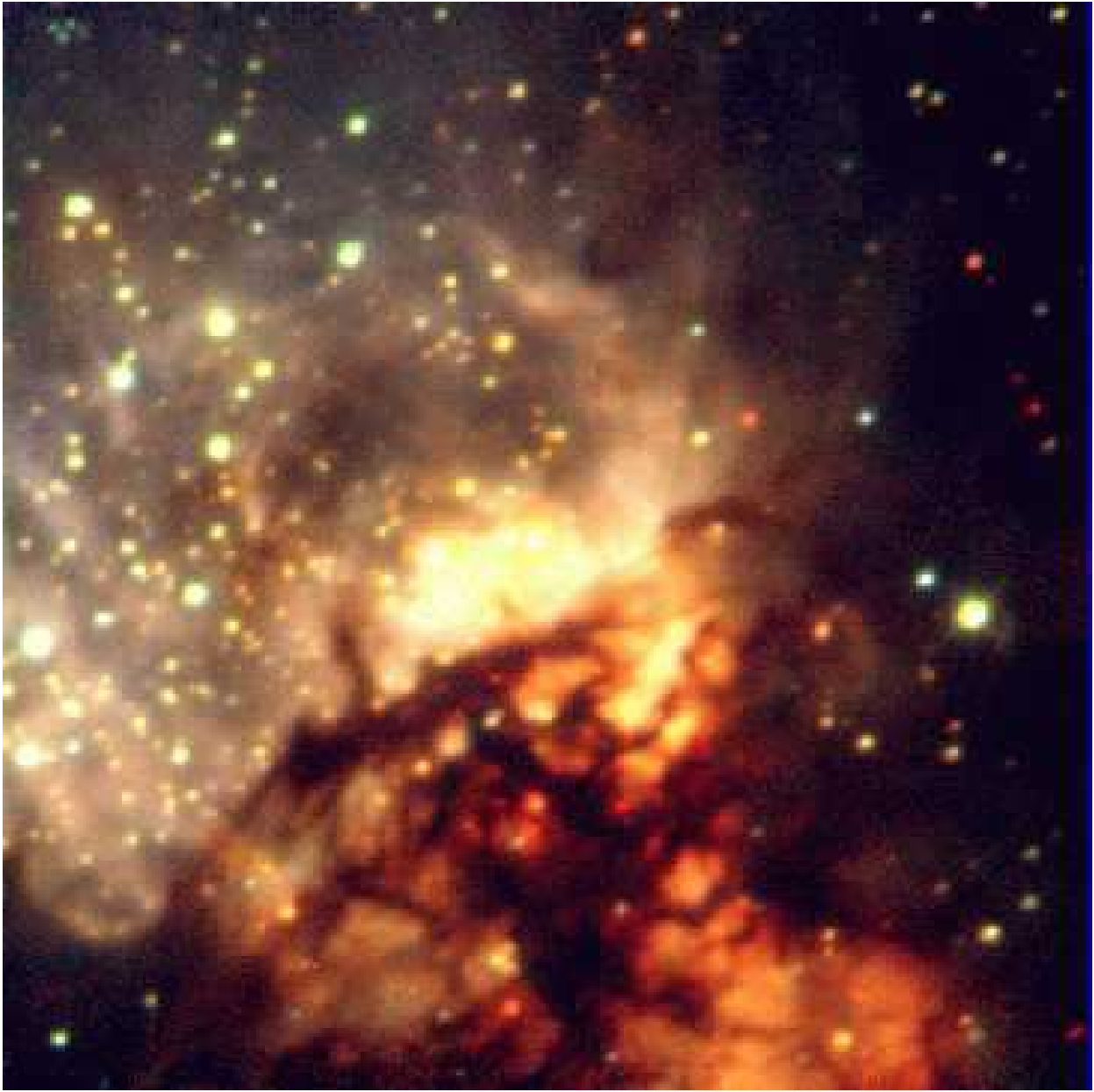}
\caption{False color image of NGC3576: $J$ is blue, $H$ is
green and $K$ is red. North is up and East to the left.\label{color}}
\end{figure}

\begin{figure}[co]
\plotone{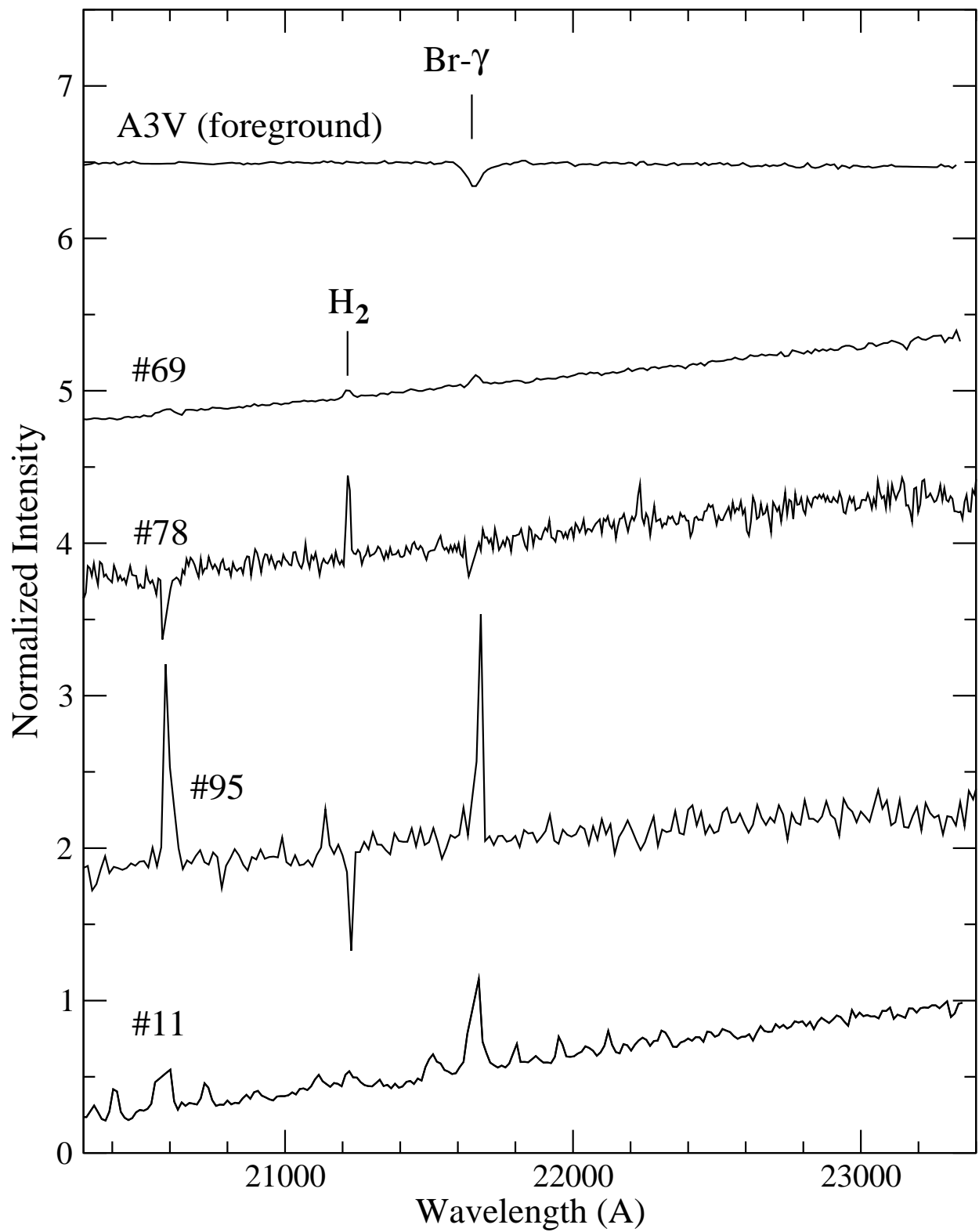}
\caption{Spectra of four cluster members displaying featureless
continuum superimposed on CO emission (\#48) or absorption (\#184,
\#180 and \#4). Star \#48 also shows H$_2$ in emission even
after ovsersubtracting He~I and Br$\gamma$ nebular component. The
spectrum at the top is a foreground M--type star. Fluxes were divided
by an A--type star and normalized to unity at 2.19 $\mu$m. 
Stars \#48 and \#160 were observed with the IRS and stars \#4, \#48 and
 \#184 with OSIRIS. \label{co}}
\end{figure}

We have presented deep $J$, $H$ and $K$ images of the newborn stellar
cluster in NGC3576 (Figure~\ref{color}) and $K-$band spectra for eight
cluster members.  The $K-$band excess emission displayed by objects
\#4, \#11 (IRS3?), \#48, \#69, \#78, \#95, \#160 and \#184, in
combination with their featureless continuum or CO emission/absorption
(Figure~\ref{co}), indicates that they are young, massive stars still
in the process of accreting material from their birth cocoons. The
lack of photospheric features and presence of disk signatures
indicates that NGC3576 is one of youngest massive star clusters in the
Milky Way.

Our data also confirm the scenario of star formation progressing from
the NE toward the inner parts of the molecular cloud (SW), but the
very young age of the cluster may contradict the claim of enhanced He
abundance in the NE part of the GH~II region by \citet{pre99}.
There are no evolved stars in the cluster which could produce the
enhancement and the time to diffuse nuclear processed material from
neighboring regions into the nebular environment would be much larger
than the cluster age. The fact that the nebular excitation increases
toward the NE, producing stronger He lines, may have influenced the
abundance calculations.
 
Since our data do not enable us to derive spectroscopic parallaxes (no
photospheric lines were detected in the luminous stars), we have
adopted the radio distance obtained by \citet{pre99} revised to
2.8 kpc by using the most recent Galactic center distance ($R_0 = 8$
kpc, \citep{r93}). The cluster parameters are not well constrained
because of distance uncertainty and the difficulty in correcting the
$K-$band magnitude for the circumstellar dust emission and
extinction. The overall picture summarized in Table~1 of a
massive and dense cluster remains valid. However, the ionizing flux
derived from the IMF is much smaller than that from radio 
observations. This is consistent with the fact that we haven't 
found spectroscopically the massive main sequence stars that 
ionize the cluster. Those stars probably remain behind heavily 
obscuring clouds. 

The fact that several of the brightest cluster members do not show
revealed photospheres raises the question: where are the ionizing
sources of NGC3576? It is plausible that some of the faint sources are
in reality luminous objects seen through large extinction, and were
not accurately dereddened because they escaped detection in the $J$
and $H-$band.  This may also explain why the number of Lyman continuum
photons derived from the IMF is smaller than that measured at radio
wavelengths.  To tackle this question, we plan to obtain spectra from
stars of $K\approx 13$ and fainter using the Gemini South 8--m
telescope.
 
In order to understand better the circumstellar environment of the
YSOs we examined here, we have performed high resolution mid-infrared
imaging with the Gemini South telescope which will be analyzed in a
future paper.  Our goal is to derive the characteristics of the
circumstellar dust, in particular evaluating its contribution to the
near infrared excess emission, extinction, and the possible presence
of accreting disks.

EF and AD thank FAPESP and PRONEX for support. PSC appreciates
continuing support from the National Science Foundation.

\clearpage


\begin{figure}[finding]
\plotone{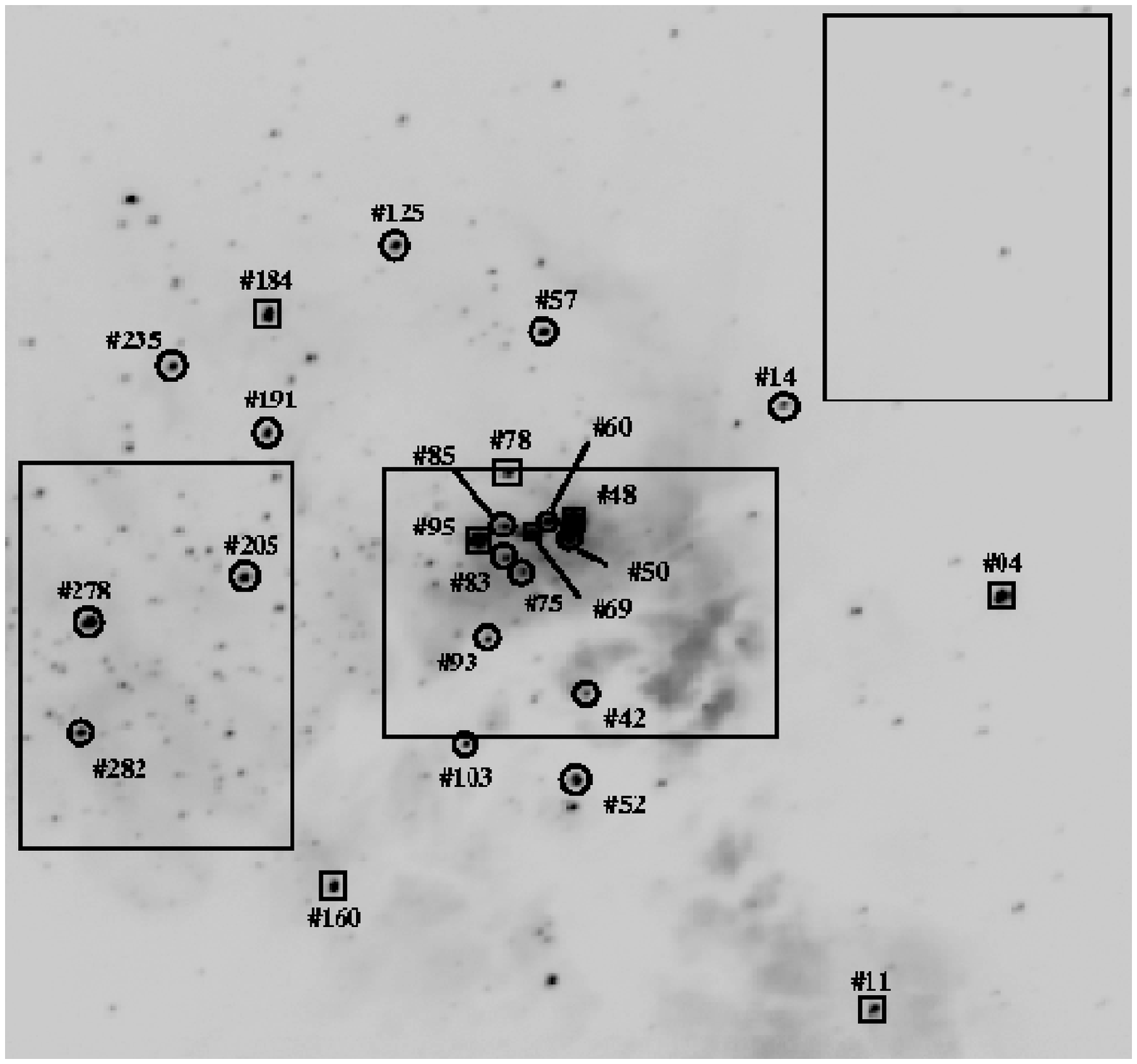}
\caption{Finding chart using a $K-$band image of NGC3576. The {\it
square symbols} are for stars whose spectra are displayed in
Figures~\ref{co} and \ref{ftless}.  North is up and East to the
left. The coordinates of the center of the image are RA (2000) = 11h11m53.6s
and Dec. = -61$^{0}$18$'$20.8$''$ and the size of the image is 
1.77$'$ x 1.65$'$. The rectangular 
boxes indicate image sub--sections used to perform artificial star experiments
(see text). \label{finding}}
\end{figure}

\clearpage 

\clearpage 

\begin{figure}[cmd]
\plotone{figueredo.fig3.eps}
\caption{$K$ vs $H - K$ color--magnitude diagram (CMD) showing the
ZAMS at $D = 2.8$ kpc and $A_K = 0.43$. The {\it open circles} indicate no
detection in the $H$ band to a limiting magnitude of $H = 15$. Object labels
are the same as in Figure 1.  \label{cmd}}
\end{figure}

\clearpage 

\clearpage 

\clearpage

\clearpage 

\clearpage 

\clearpage

\clearpage

\clearpage

\clearpage

\pagestyle{empty}
\begin{deluxetable}{ll}
\tablecaption{Cluster Properties\label{properties}} \tablewidth{0pt}
 \tablehead{ }
 \startdata
Cluster distance & 2.8 $\pm$ 0.3 kpc\\
Cluster diameter& 1.5 pc\\
Cluster mass& $<$5.4 x 10$^3$ M$_\odot$\\
Stellar density& $<$3.1 x 10$^3$ M$_\odot$ pc$^{-3}$\\
IMF slope & $\Gamma = - 1.62$\\
NLyc phot. (IMF)& 0.42 - 1.67 x 10$^{50}$s$^{-1}$\\
NLyc phot. (radio)& 1.6 $\pm$ 0.4 x 10$^{50}$s$^{-1}$\\
\enddata  

\end{deluxetable}

\pagestyle{empty}
\begin{deluxetable}{lrcrrl}
\tablecaption{YSO properties \label{YSOs}} \tablewidth{0pt}
 \tablehead{ \colhead{ID} & \colhead{$J-H$\tablenotemark{a}} &
\colhead{$H-K$\tablenotemark{a} } & \colhead{$K$\tablenotemark{a}} &
K-exc\tablenotemark{b} &  \colhead{Note}}
 \startdata
 \#4   &1.51   &0.93& 9.95 & -0.01 &CO abs\\
 \#11  &$>$0.99  &1.10& 12.91& -0.26 &\\
\#48  &3.04   &2.21& 8.35 & -3.17 & H$_2$ \& CO em\\
\#69  &$>$3.44  &1.45&10.10 & -0.62 &H$_2$ em\\
\#78  &2.02   &1.38&11.67 & -0.16 &H$_2$ em\\
\#95  &1.67   &1.41& 9.26 & -0.40 &\\
\#160 &$>$0.95  &2.93& 11.12& -2.09 &CO abs\\
\#184 &1.23   &0.81&10.40 & -0.07 &CO abs\\
\enddata \tablenotetext{a}{Non-derredned magnitudes; errors 
($J-H\pm0.05$, $H-K\pm0.08$, $K\pm0.07$)
}\tablenotetext{b}{Excess magnitude in the $K-$band after derredening} 

\end{deluxetable}

\end{document}